\documentclass[%
 aip,
 amsmath,amssymb,
 reprint,%
]{revtex4-2}

\usepackage{graphicx}
\usepackage{dcolumn}
\usepackage{bm}
\usepackage[breaklinks=true,colorlinks=true,linkcolor=blue,urlcolor=blue,citecolor=blue]{hyperref}

\usepackage[utf8]{inputenc}
\usepackage[T1]{fontenc}
\usepackage{mathptmx}
\usepackage{etoolbox}
\usepackage{xr}
\makeatletter
\def\@email#1#2{%
 \endgroup
 \patchcmd{\titleblock@produce}
  {\frontmatter@RRAPformat}
  {\frontmatter@RRAPformat{\produce@RRAP{*#1\href{mailto:#2}{#2}}}\frontmatter@RRAPformat}
  {}{}
}%
\makeatother

\begin{document}

\title{Site-selective enhancement of Eu emission in delta-doped GaN}

\author{Amelia R. Klein}
\affiliation{Quantum Engineering Laboratory, Department of Electrical and Systems Engineering, University of Pennsylvania, 200 S. 33rd St. Philadelphia, Pennsylvania, 19104, USA}

\author{Hayley J. Austin}
\affiliation{Department of Physics, Lehigh University, Bethlehem, Pennsylvania 18015, USA}

\author{Fumikazu Murakami}
\affiliation{Institute of Laser Engineering, Osaka University, 2-6 Yamada-oka, Suita, Osaka 565-0871, Japan}

\author{Jamie Ford}
\affiliation{Singh Center for Nanotechnology, University of Pennsylvania, Philadelphia, Pennsylvania, 19104, USA}

\author{Jun Tatebayashi}
\affiliation{Department of Materials and Manufacturing Science, Graduate School of Engineering, Osaka University, 2-1 Yamada-oka, Suita, Osaka 565-0871, Japan}

\author{Masayoshi Tonouchi}
\affiliation{Institute of Laser Engineering, Osaka University, 2-6 Yamada-oka, Suita, Osaka 565-0871, Japan}

\author{Yasufumi Fujiwara}
\affiliation{Intra-Photonics Research Center, Research Organization of Science and Technology, Ritsumeikan University, 1-1-1 Nojihigashi, Kusatsu, Shiga 525-8577, Japan}
\affiliation{Department of Materials and Manufacturing Science, Graduate School of Engineering, Osaka University, 2-1 Yamada-oka, Suita, Osaka 565-0871, Japan}

\author{Volkmar Dierolf}
\affiliation{Department of Physics, Lehigh University, Bethlehem, Pennsylvania 18015, USA}

\author{Lee C. Bassett}
\affiliation{Quantum Engineering Laboratory, Department of Electrical and Systems Engineering, University of Pennsylvania, 200 S. 33rd St. Philadelphia, Pennsylvania, 19104, USA}

\author{Brandon Mitchell}
\affiliation{Department of Physics, West Chester University, West Chester, Pennsylvania 19383, USA}
\affiliation{Department of Physics, Lehigh University, Bethlehem, Pennsylvania 18015, USA}
\affiliation{Department of Materials and Manufacturing Science, Graduate School of Engineering, Osaka University, 2-1 Yamada-oka, Suita, Osaka 565-0871, Japan}
\email[Corresponding authors. ]{bmitchell@wcupa.edu \& lbassett@seas.upenn.edu}

\begin{abstract}
Europium-doped gallium nitride (GaN:Eu) is a promising platform for classical and quantum optoelectronic applications.
When grown using organometallic vapor-phase epitaxy, the dominant red emission from Eu exhibits an inhomogeneous photoluminescence (PL) spectrum due to contributions from several non-equivalent incorporation sites that can be distinguished with combined excitation emission spectroscopy.
Energy transfer from the GaN bandgap to the majority site is inefficient, limiting the performance of GaN:Eu LEDs and resulting in an inhomogeneous emission spectrum dominated by disproportionate contributions from minority sites.
In this work, we use site-selective spectroscopy to characterize the photoluminescence properties of delta-doped structures with alternating doped and undoped layers of varying thicknesses and demonstrate that they selectively enhance emission from the majority site when compared to uniformly-doped samples.
Samples with 2-nm and 10-nm doped layers show much greater PL intensity per Eu concentration as well as more efficient energy transfer to the majority site, which are both highly desirable for creating power-efficient LEDs.
Meanwhile, a sample with 1-nm doped layers shows emission only from the majority site, resulting in a narrow, homogeneous emission spectrum that is desirable for quantum technologies.
This utilization of delta-doping has the potential to be broadly applicable for engineering desirable defect properties in rare-earth doped semiconductors.
\end{abstract}

\maketitle
\UseRawInputEncoding

\begin{figure*}
    \includegraphics{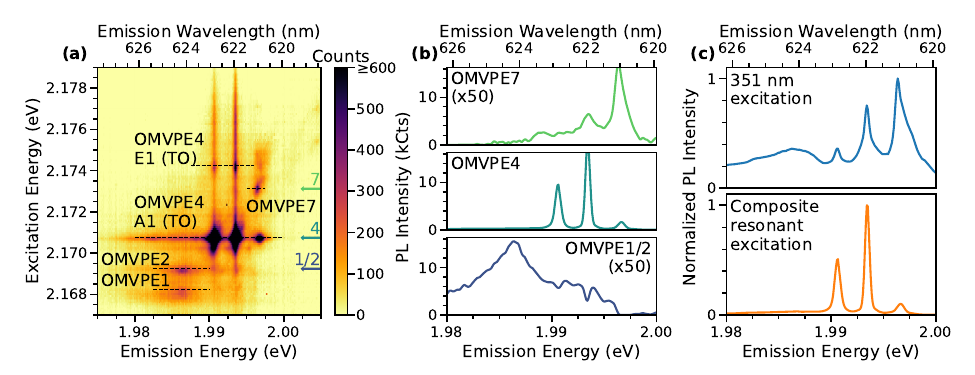}
    \caption{Europium incorporation sites in uniformly-doped GaN.
    \textbf{(a)} Combined excitation-emission spectroscopy (CEES) of a uniformly doped GaN:Eu sample showing distinct incorporation sites labeled with dashed lines to indicate relevant regions.
    Two distinct peaks in excitation are shown for OMVPE4 due to coupling to different phonons.
    \textbf{(b)} Emission spectra of the three dominant sites obtained by spectral decomposition of the CEES map.
    Extracted spectra are shown at their peak amplitudes at energies noted by colored arrows in \textbf{(a)}.
    \textbf{(c)}Composite emission spectrum obtained by summing the spectra in \textbf{(b)} (bottom, orange) compared to measured emission spectrum of the same sample when excited above bandgap with a 351 nm laser (top, blue).}
    \label{fig:sites}
\end{figure*}

Europium-doped GaN (GaN:Eu) is a promising platform for displays and quantum technologies.
It exhibits bright, narrow-bandwidth red emission centered around 622 nm resulting from the $~^5\!D_0 \rightarrow ~^7\!F_2$ transition of the trivalent $\mathrm{Eu^{3+}}$ ion, which can be pumped electrically or optically.
Red LEDs based on GaN:Eu have shown external quantum efficiencies up to 9.1\% \cite{Mitchell2018Perspective:Doping} and complement existing high-efficiency blue and green LEDs on GaN \cite{Yu2022Ultra-smallLithography, Pandey2023AnMicro-LED}, allowing for the realization of high-density monolithically stacked full-color LEDs \cite{Ichikawa2021Eu-dopedGamut}.
Additionally, color tunability of GaN:Eu LEDs has been demonstrated by stimulating emission from the $~^5\!D_1$ level \cite{Zhu2018Re-ExcitationExcitation, Austin2022ModelingLayers, Mitchell2019Color-TunablilityInjection}.

Rare-earth ions such as $\mathrm{Eu^{3+}}$ are also promising for emerging quantum information technologies due to their long spin coherence times and narrow optical linewidths resulting from the shielding of their 4f shell electrons \cite{Nilsson2002InitialCrystals, Thiel2011Rare-earth-dopedProcessing, Ourari2023IndistinguishableState, Uysal2025Spin-PhotonBand, Serrano2022Ultra-narrowCrystals, Zhong2019EmergingNanophotonics, Guo2023Rare-earthQuo, Pettit2023AIons}.
Eu ions in particular exhibit exceptionally long nuclear spin coherence times due to their spin-zero electronic ground state \cite{Konz2003Temperature5, Yano1991UltralongEu3+:Y2SiO5, Arcangeli2014Spectroscopy5, Zhong2015OpticallyTime, Ma2023MonteCrystals}, enabling demonstrations of long-lived quantum memories \cite{Jobez2015CoherentMemory, Ma2021One-hourMemory}.
Unlike typical complex oxide hosts, GaN is a technologically mature wide-bandgap semiconductor platform, facilitating easy device integration with nanoelectronics and nanophotonics, such as photonic crystal cavities for enhancing GaN:Eu emission \cite{Ichikawa2021EnhancedHoles, Iwaya2022ImprovedEngineering}.
Additionally, GaN's band gap ($\sim$364 nm) is accessible with near-UV excitation, with efficient indirect electrical or optical excitation of $\mathrm{Eu^{3+}}$ ions enabled by fast ($<100$ ps) energy transfer from excited charge carriers \cite{Wei2019PicosecondIons, Timmerman2020CarrierGaN}.

The development of GaN:Eu---typically grown using organometallic vapor-phase epitaxy (OMVPE)---toward both classical and quantum technologies is hindered by the formation of non-equivalent incorporation sites.
The local environments and symmetries of these $\mathrm{Eu^{3+}}$ sites results in distinct excitation energies and emission spectra \cite{Binnemans2015InterpretationSpectra, Fleischman2009ExcitationSpectroscopy, Woodward2011SiteLayers, Wakamatsu2013LuminescenceSites}, as revealed by combined excitation-emission spectroscopy (CEES) in Fig.~\ref{fig:sites}a.
The sites are named OMVPE1-8 after the synthesis method and are ordered by their excitation energies at the zero-phonon transition of $~^7\!F_0 \rightarrow ~^5\!D_0$ ($\sim$589 nm) \cite{Woodward2011SiteLayers}.
Fig.~\ref{fig:sites}a shows the phonon-assisted excitations of this transition near 571 nm (2.17 eV).
The phonon-assisted excitation relaxes the forbidden $\Delta J = 0$ selection rule so that its efficiency is not strongly dependent on site symmetry, allowing for the brightness of each site to serve as a good approximation of its relative abundance \cite{Woodward2011SiteLayers, Copelman2020StrongGaN}.

The $~^5\!D_0 \rightarrow ~^7\!F_2$ emission spectra of the most prevalent sites are extracted from the CEES map and plotted in Fig.~\ref{fig:sites}b.
These spectra and their amplitudes were obtained from a spectral decomposition of the CEES data using non-negative matrix factorization.
The emission under resonant excitation in CEES is dominated by the majority site, OMVPE4, which  composes $>90\%$ of $\mathrm{Eu^{3+}}$ ions in this sample.
In contrast, the emission spectrum under above-bandgap excitation (e.g., at 351~nm as shown in Fig.~\ref{fig:sites}c) contains disproportionate contributions from OMVPE1/2 and from OMVPE7 (also known as Eu-2) \cite{Woodward2011SiteLayers, Woodward2011ExcitationCenter, Wakamatsu2013LuminescenceSites}.
This is a key signature of inefficient energy transfer to the majority site, which limits the external quantum efficiency of GaN:Eu LEDs \cite{Mitchell2018Perspective:Doping} and results in an inhomogeneously broadened emission spectrum that is undesirable for quantum applications.
Although past work has considered increasing the formation of the more efficient minority sites \cite{Iwaya2023EnhancedAnnealing}, here we focus on increasing the efficiency of the majority site.

One method of improving the excitation efficiency of rare-earth defects is through the formation of superlattice structures.
For example, GaN/AlN multiple quantum wells (MQWs) have been shown to increase the emission efficiency of GaN:Er by an order of magnitude \cite{Altahtamouni2015DramaticStructures, Ho2018Room-TemperatureRegion}.
Similar enhancement has also been observed in Si:Er systems using alternating Si/Si:Er layers \cite{Stepikhova2001Properties320323, Vinh2003MicroscopicSilicon, Vinh2004OpticalNanostructures, Vinh2007ConcentrationNanolayers, Kuznetsov2000SpecialStructures}.
In GaN:Eu specifically, it was previously demonstrated that LEDs consisting of alternating GaN/GaN:Eu layers exhibit a factor of 25 increase in output power over structures with a single active layer \cite{Zhu2017High-PowerEpitaxy, Mitchell2018Perspective:Doping}.
The leading explanation for these observations is that shallow wells between doped and undoped layers induce carrier confinement in the doped regions; this is supported by terahertz spectroscopy experiments showing that dilute ($\sim$0.1\%) Eu doping leads to a $\sim$40 meV reduction in the GaN bandgap \cite{Murakami2023EnhancedSpectroscopy}.

\begin{figure*}
    \includegraphics{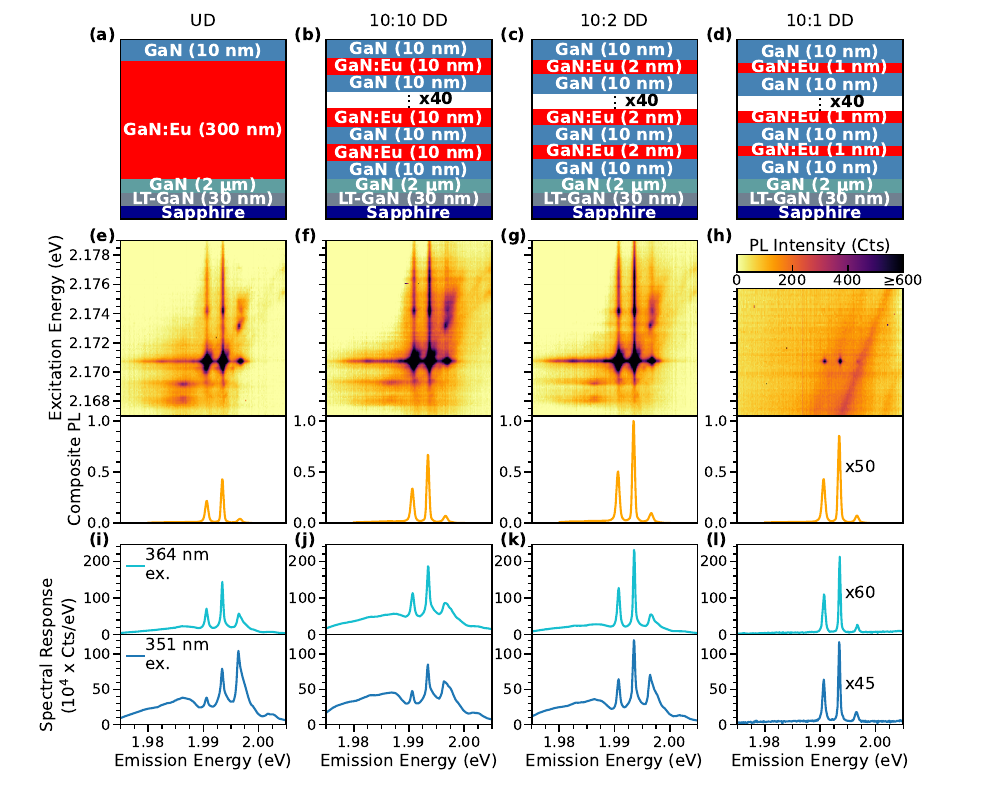}
    \caption{Site-selective spectroscopy of uniformly doped (UD) and delta-doped (DD) GaN:Eu.
    \textbf{(a)}-\textbf{(d)} Cross-sectional schematics of GaN:Eu samples studied in this work.
    All samples are grown on a c-plane sapphire substrate with a 30-nm-thick low-temperature (LT) buffer layer and a 2-µm-thick undoped GaN layer before the Eu-doped region.
    \textbf{(a)} is the same uniformly-doped sample depicted in Fig.~\ref{fig:sites} with a 300nm thick doped region, while \textbf{(b)}-\textbf{(d)} contain 40 pairs of alternating doped and undoped layers of different thicknesses.
    \textbf{(e)}-\textbf{(h)} (Above) Resonant CEES maps, plotted using a shared colormap with thresholded counts such that the minority sites are clearly visible.
    (Below) Composite emission spectra calculated by summing the OMVPE1/2, OMVPE4, and OMVPE7 components obtained via spectral decomposition.
    Amplitudes are normalized to the peak of the brightest sample.
    \textbf{(i)}-\textbf{(l)} Emission spectra of each sample taken with near-UV excitation at 351 nm (above bandgap, lower row) and 364 nm (below bandgap, upper row).}
    \label{fig:megaplot}
\end{figure*}

In this work, we study four GaN:Eu samples: one uniformly doped (UD) sample with a 300-nm-thick GaN:Eu active layer (Fig.~\ref{fig:megaplot}a) and three delta-doped samples, labeled ``$a$:$b$ DD,'' consisting of 40 alternating pairs of $a$-nm-thick GaN and $b$-nm-thick GaN:Eu (Fig.~\ref{fig:megaplot}b-d).
GaN:Eu layers were grown using OMVPE at 960 \textdegree C using an oxygen-free liquid precursor ($\mathrm{EuCp^{pm}_2}$) \cite{Mitchell2017SynthesisEpitaxy} and are co-doped with oxygen to facilitate stable Eu incorporation \cite{Mitchell2016UtilizationApplications}.
Further details of the sample growth have been reported elsewhere \cite{Mitchell2018Perspective:Doping, Ichikawa2021Eu-dopedGamut, Zhu2016EnhancedEnvironment}, and additional growth parameters are tabulated in Table I in the Supplementary Information.

CEES measurements of these samples shown in Fig.~\ref{fig:megaplot}e-h were performed at 10K in a closed-cycle cryostat (Montana Instruments) using a tunable Model 590 dye laser.
The data from Fig.~\ref{fig:sites} for the UD sample are reproduced in Figs. \ref{fig:megaplot}e and \ref{fig:megaplot}i for easy comparison with DD samples.
Notably, in the 10:1 DD sample, we observe emission only from OMVPE4, albeit with a dramatically lower intensity compared to the other samples.
Emission spectra of these four samples taken using UV excitation just above the GaN bandgap (351 nm) and just below bandgap (364 nm) are plotted in Fig.~\ref{fig:megaplot}i-l.
Above-bandgap excitation generally results in an enhanced emission peak at 1.997 eV associated with OMVPE7, whereas below-bandgap excitation transfers energy more efficiently to OMVPE4 and therefore produces a stronger central peak at 1.994 eV \cite{Woodward2011ExcitationCenter}.

As the thickness of the GaN:Eu layers decreases from the UD sample to the 10:10, 10:2, and finally 10:1 DD structures, the relative contribution of OMVPE4 to the emission spectra increases, as evidenced by the relative peak heights.
A quantitative analysis of fitting the UV emission spectra to the spectral components in Fig.~\ref{fig:sites}b bears this out, as described in the Supplementary Information (Figs. S3-S4).
Unlike the other samples, the 10:1 DD sample has identical emission spectra under above-bandgap, below-bandgap, and resonant excitation; this is consistent with the observation that only OMVPE4 exists in detectable quantities.

It is further notable from Fig.~\ref{fig:megaplot} that the 10:2 DD sample is brighter than the UD and 10:10 DD samples, despite the total thickness of its GaN:Eu layers being significantly smaller.
An increase in the efficiency of the energy transfer to OMVPE4, consistent with previous observations with LEDs \cite{Mitchell2018Perspective:Doping}, would explain the enhanced contribution from OMVPE4 in the the UV-excited spectra. 
Meanwhile, the increased brightness seen under phonon-assisted resonant excitation (see the composite emission spectra in Fig.~\ref{fig:megaplot}e-h) suggests either an unexpectedly high Eu concentration or a contradiction of the assumption that the intensity of the phonon-assisted excitation is proportional to the site concentration.

\begin{figure*}
    \includegraphics{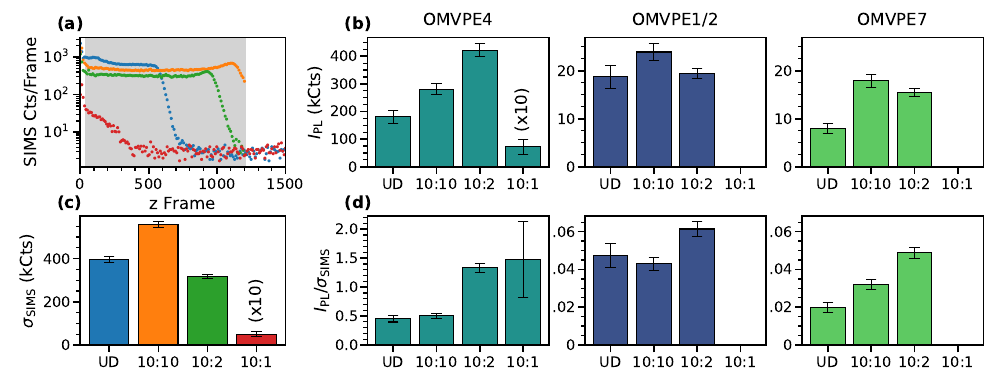}
    \caption{Eu concentration and resonant brightness of GaN:Eu samples.
    \textbf{(a)} SIMS counts per frame detected for $\mathrm{~^{151}Eu}$ over a 50 $\mu$m field of view as a function of image frame, rebinned every 10 frames to reduce noise.
    Each frame corresponds to an etched depth of 0.5-0.6 nm.
    \textbf{(b)} Integrated PL brightness ($I_{\mathrm{PL}}$) of OMVPE2, OMVPE4, and OMVPE7 extracted from CEES maps using spectral decomposition.
    Values are rescaled according to the measured absolute brightness of OMVPE4 of each sample, with error bars corresponding to the standard deviation of counts over a spatially-resolved photoluminescence scan.
    \textbf{(c)} Total number of SIMS counts ($\sigma_{\mathrm{SIMS}}$) for each sample, taken by summing over the shaded region in \textbf{(a)}.
    Error bars correspond to the standard deviation of rebinned data points.
    \textbf{(d)} Concentration-normalized photoluminescence intensity, $I_{\mathrm{PL}}/\sigma_{\mathrm{SIMS}}$.}
    \label{fig:resonant_intensity}
\end{figure*}

In order to more directly compare the sample brightnesses with their Eu concentrations while accounting for inconsistencies during growth and sample preparation, we performed time-of-flight secondary ion mass spectrometry (ToF-SIMS) measurements on each sample using a focused ion beam scanning electron microscope (TESCAN S8000X) on a cryogenic stage cooled to $-$50 \textdegree C.
Figure~\ref{fig:resonant_intensity}a shows the total $\mathrm{~^{151}Eu}$ counts per SIMS measurement frame (each frame is $\sim$0.5-0.6 nm of depth and a 50 \textmu m square area), and Fig.~\ref{fig:resonant_intensity}c shows the total counts summed over the volume, $\sigma_{\mathrm{SIMS}}$.
Modulation of the $\mathrm{~^{151}Eu}$ counts is not clearly visible in Fig.~\ref{fig:resonant_intensity}a due to a low signal-to-noise ratio and area averaging, but an additional room-temperature SIMS measurement included in the Supplementary Information confirms the presence of distinct GaN/GaN:Eu layers (Fig.~S1).
Resonant photoluminescence (PL) intensities ($I_{\mathrm{PL}}$) of each site and sample shown in Fig.~\ref{fig:resonant_intensity}b were extracted from the CEES maps in Fig.~\ref{fig:megaplot}e-h using spectral decomposition; their amplitudes were additionally corrected using a second set of calibration measurements to ensure accurate comparisons of absolute intensity across samples.
Finally, Fig.~\ref{fig:resonant_intensity}d shows the PL intensity per $\mathrm{~^{151}Eu}$ count, $I_{\mathrm{PL}}/\sigma_{\mathrm{SIMS}}$, which represents the concentration-normalized intensity of each incorporation site.
Further details of the spectral decomposition, amplitude correction, and SIMS measurements are included in the Supplementary Information.

In the SIMS depth profiles (Fig.~\ref{fig:resonant_intensity}a), the UD, 10:10 DD, and 10:2 DD samples all show a roughly constant $\mathrm{~^{151}Eu}$ count rate from the surface to a depth  consistent with the samples' layer thicknesses (300 nm for UD, 800 nm for 10:10 DD, and 480 nm for 10:2 DD, with each frame corresponding to 0.5-0.6 nm).
The 10:1 DD sample, on the other hand, has a dramatically lower $\mathrm{~^{151}Eu}$ count rate near the surface, and the Eu signal does not extend as deep as expected (around 440 nm).
It is possible that the particularly thin active layers of the 10:1 DD sample ($\sim$2 GaN lattice constants) inhibit the stable incorporation of Eu into the lattice, resulting in a lower Eu concentration overall. 
This would also explain the absence of minority sites that depend on more elaborate complexes of nearest-neighbor Eu ions and trap states.
The decreased Eu incorporation and doping depth observed in the 10:1 DD sample might also result from the radial position of this particular sample on the full growth wafer; a systematic analysis of Eu doping as a function of wafer location would be illustrative to distinguish these possibilities.

From the concentration-normalized intensities (Fig.~\ref{fig:resonant_intensity}d), the 10:2 DD and 10:1 DD samples show a nearly factor of 3 enhancement of OMVPE4 intensity compared to the UD and 10:10 DD samples.
Similar enhancement is also observed for OMVPE7 in the 10:10 DD and 10:2 DD samples, whereas no clear trend is observed for OMVPE1/2.

The simultaneous enhancement of multiple sites indicates that the resonant PL enhancement is not a result of preferential site formation; however, it could indicate that a greater proportion of Eu ions are optically active.
The resonant intensity enhancement is also not due to a change in the $~^5\!D_0$ excited state lifetime; measured OMVPE4 lifetimes (plotted in Fig. S6 in the Supplementary Information) are nearly identical (272-274 $\mu$s) for the UD, 10:10 DD, and 10:2 DD samples and slightly increased ($306.0 \pm 0.9 \mu$s) for the 10:1 DD sample.
Intensity enhancement could still be explained by quenching of non-radiative decay pathways, increased coupling to the phonon mode used in excitation, or some photonic effect (doped layers exhibit a $\sim$20\% refractive index increase \cite{Murakami2023EnhancedSpectroscopy}) influencing the excitation or collection efficiency.

\begin{figure*}
    \includegraphics{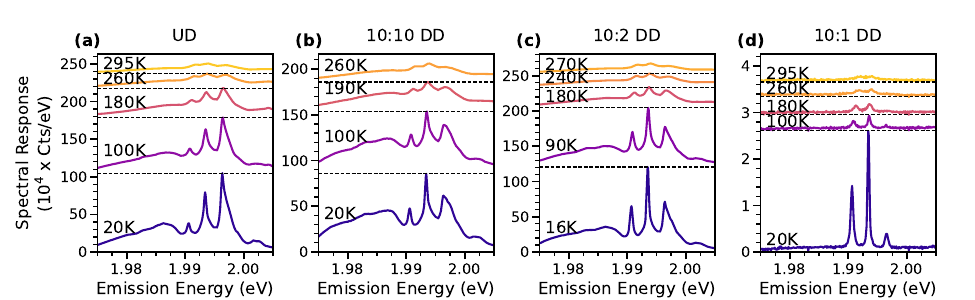}
    \caption{Temperature dependence of the emission spectra of the four samples taken with above-bandgap (351 nm) excitation, with dashed lines indicating the effective zeros for each curve.}
    \label{fig:temperature}
\end{figure*}

\begin{figure}
    \includegraphics{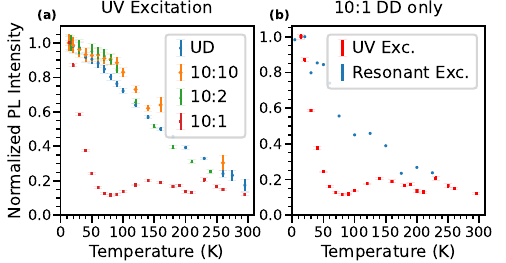}
    \caption{Temperature dependence of the total integrated intensity of the $~^5\!D_0 \rightarrow ~^7\!F_2$ emission.
    For each sample, the data are normalized to the intensity at low temperature.
    \textbf{(a)} compares the total intensity for all four samples under above-bandgap (351 nm) excitation.
    \textbf{(b)} compares the total intensity of the 10:1 DD sample under above bandgap excitation (red) and under phonon-assisted $~^7\!F_0 \rightarrow ~^5\!D_0$ excitation (blue).}
    \label{fig:intensity_vs_temperature}
\end{figure}

In order to further understand the enhancement of energy transfer under indirect excitation, we measured the temperature dependence of emission spectra under above-bandgap (351 nm) excitation; see Fig.~\ref{fig:temperature} for spectra at representative temperatures between 16 - 295~K.
In every case as the temperature increases, the emission spectra broaden, the peaks become less resolvable, and the overall intensity decreases.
However, subtle details in the spectral shape and amplitude reveal important differences between the samples.

Fig.~\ref{fig:intensity_vs_temperature}a shows the normalized integrated intensity for each sample versus temperature.
The intensity of the 10:1 DD sample drops dramatically between 20-50K before leveling off, while the intensities of the 10:10 DD and 10:2 DD samples remain more stable over 0-100K than the UD sample, where the intensity falls consistently with increasing temperature.
These observations are consistent with the hypothesis that the bandgap reduction associated with doped layers \cite{Murakami2023EnhancedSpectroscopy} results in carrier trapping that increases the efficiency of energy transfer to $\mathrm{Eu^{3+}}$ sites.
Such carrier trapping is expected to be effective when the depth of the well exceeds the available thermal energy and hence manifests as a decrease in excitation efficiency above a characteristic temperature threshold.
In the case of the 10:1 DD sample, the reduced concentration and thin layers results in a particularly shallow well and correspondingly low temperature threshold; meanwhile, the 10:10 DD and 10:2 DD samples exhibit enhancement to higher temperatures compared to the UD sample, before their intensities drop around 100K.

Separately, the intensity of the 10:1 DD sample actually increases with temperature between 90-140K; this feature is not observed in the other samples.
Similar intensity increases around 140K have been observed in other GaN:Eu incorporation sites \cite{Katchkanov2005PhotoluminescenceEpilayers, Wakamatsu2014AfterglowEpitaxy, Iwaya2024AnGaN} and are likely due to the release of carriers from the ionization of shallow GaN trap states \cite{Reshchikov2018ThermalGaN, Reshchikov2020GiantGaN}, though these intensity fluctuations have not previously been seen for OMVPE4.
Both the sharp decrease in temperature from 20-50K and the increase around 140K appear only under indirect excitation, as shown in Fig.~\ref{fig:intensity_vs_temperature}b, providing direct evidence that these phenomena are related to carrier dynamics and energy-transfer rather than the radiative efficiency of OMVPE4.

In conclusion, we have demonstrated multiple forms of enhancement of red $~^5\!D_0 \rightarrow ~^7\!F_2$ emission in delta-doped GaN/GaN:Eu layered samples.
First, we observe improved energy transfer to the majority site from the shape of the emission spectra; this is consistent with previous measurements of the quantum efficiency of GaN:Eu LEDs \cite{Mitchell2018Perspective:Doping}.
Meanwhile, the changes in thermal quenching behavior of DD samples support the prediction of carrier trapping in the doped layers due to a bandgap reduction \cite{Murakami2023EnhancedSpectroscopy}.
Unexpectedly, we also observe enhancement of the phonon-assisted resonant emission, which may indicate an improvement in optically-active Eu incorporation.
All of these enhancements are directly beneficial in improving the efficiency of GaN:Eu optoelectronics.

At the same time, we also demonstrate a GaN:Eu sample that features only a single $\mathrm{Eu^{3+}}$ incorporation site in detectable quantities.
Although its overall brightness is lower than the other samples, the pure OMVPE4 emission is still enhanced when considering the concentration-normalized brightness.
Its narrow, homogeneous emission spectra is highly desirable for applications in quantum technologies.
Moreover, the invariance of its emission spectrum across multiple forms of indirect excitation (above-bandgap and just-below-bandgap) and resonant excitation suggests the possibility of probing or initializing $\mathrm{Eu^{3+}}$ ensembles using multiple techniques.
While other studies have reported a single optical site---albeit with a broad (several nm) excitation line---by codoping GaN:Eu with Mg\cite{Sekiguchi2019ObservationEpitaxy}, this method of delta-doping does not require any additional dopants or materials.

The particularly bright 10:2 DD sample and the dim but homogeneous 10:1 DD sample are interesting for different application domains, yet both were grown using the same delta-doping technique simply by tuning the doped layer thicknesses.
Optimizing the doping concentrations and undoped layer thicknesses in conjunction with the doped layer thicknesses can further extend the ability to achieve both homogeneity and overall brightness.
In tandem with these favorable site properties, the refractive index contrast between layers could be utilized to engineer desirable optical properties.
Because of its simplicity and tunability, and supported by similar observations in Si:Er \cite{Stepikhova2001Properties320323, Vinh2003MicroscopicSilicon, Vinh2004OpticalNanostructures, Vinh2007ConcentrationNanolayers, Kuznetsov2000SpecialStructures}, delta-doping techniques could be useful for achieving desired material characteristics for various combinations of rare-earth ions and semiconductor hosts.

\bigbreak

See the supplementary material for further details of the sample growth, measurements, and spectral analysis, as well as excited-state lifetime measurements and an additional room-temperature SIMS measurement that more clearly resolves the doped and undoped layers.

\bigbreak

This work was supported by the NSF under award No. ECCS-2129183. 
This work was carried out in part at the Singh Center for Nanotechnology, which is supported by the NSF National Nanotechnology Coordinated Infrastructure Program under grant NNCI-2025608, and through the use of facilities supported by the University of Pennsylvania Materials Research Science and Engineering Center (MRSEC) DMR-2309043.
This work was partially supported by JSPS KAKENHI (Grant No. 23H05449) and JST D-Global (Grant No. 24015395).

\section*{Author Declarations}

\subsection*{Conflict of Interest}
The authors declare the following potential conflict of interest: one or more authors are inventors on a patent application related to the subject matter of this article (Patent Application No. [US63/930,461]), which has been submitted and is currently pending.

\subsection*{Author Contributions}
\textbf{Amelia R. Klein}: Formal Analysis (lead); Investigation (equal); Validation (lead); Visualization (lead); Writing -- Original Draft (lead).
\textbf{Hayley J. Austin}: Investigation (equal); Formal Analysis (supporting); Writing -- Original Draft (supporting).
\textbf{Fumikazu Murakami}: Investigation (equal); Writing -- Original Draft (supporting).
\textbf{Jamie Ford}: Investigation (supporting); Resources (equal).
\textbf{Jun Tatebayashi}: Resources (equal).
\textbf{Masayoshi Tonouchi}: Resources (equal).
\textbf{Yasufumi Fujiwara}: Conceptualization (equal); Resources (equal).
\textbf{Volkmar Dierolf}: Conceptualization (equal); Funding acquisition (equal); Supervision (equal).
\textbf{Lee C. Bassett}: Funding acquisition (equal); Supervision (equal); Writing -- Review and Editing (equal).
\textbf{Brandon Mitchell}: Conceptualization (equal); Funding acquisition (equal); Project Administration (lead); Supervision (equal); Writing -- Review and Editing (equal).

\section*{Data Availability}
The data that support the findings of this study are available from the corresponding author upon reasonable request.

\bibliography{references}

@article{Pettit2023AIons,
    title = {{A perspective on the pathway to a scalable quantum internet using rare-earth ions}},
    year = {2023},
    journal = {Applied Physics Reviews},
    author = {Pettit, Robert M and Farshi, Farhang Hadad and Sullivan, Sean E and V{\'{e}}liz-Osorio, Álvaro and Singh, Manish Kumar},
    number = {3},
    month = {8},
    pages = {031307},
    volume = {10},
    url = {https://doi.org/10.1063/5.0156874},
    doi = {10.1063/5.0156874},
    issn = {1931-9401}
}

@article{Wakamatsu2014AfterglowEpitaxy,
    title = {{Afterglow of Eu-related emission in Eu-doped gallium nitride grown by organometallic vapor phase epitaxy}},
    year = {2014},
    journal = {Journal of Applied Physics},
    author = {Wakamatsu, R and Timmerman, D and Lee, D and Koizumi, A and Fujiwara, Y},
    number = {4},
    month = {7},
    pages = {043515},
    volume = {116},
    url = {https://doi.org/10.1063/1.4891232},
    doi = {10.1063/1.4891232},
    issn = {0021-8979}
}

@article{Iwaya2024AnGaN,
    title = {{An efficiently excited Eu3+ luminescent site formed in Eu,O-codoped GaN}},
    year = {2024},
    journal = {AIP Advances},
    author = {Iwaya, Takenori and Ichikawa, Shuhei and Dierolf, Volkmar and Mitchell, Brandon and Austin, Hayley and Timmerman, Dolf and Tatebayashi, Jun and Fujiwara, Yasufumi},
    number = {2},
    month = {2},
    pages = {25044},
    volume = {14},
    url = {https://doi.org/10.1063/5.0183774},
    doi = {10.1063/5.0183774},
    issn = {2158-3226}
}

@article{Pandey2023AnMicro-LED,
    title = {{An Ultrahigh Efficiency Excitonic Micro-LED}},
    year = {2023},
    journal = {Nano Letters},
    author = {Pandey, Ayush and Min, Jungwook and Reddeppa, Maddaka and Malhotra, Yakshita and Xiao, Yixin and Wu, Yuanpeng and Sun, Kai and Mi, Zetian},
    number = {5},
    month = {3},
    pages = {1680--1687},
    volume = {23},
    publisher = {American Chemical Society},
    url = {https://doi.org/10.1021/acs.nanolett.2c04220},
    doi = {10.1021/acs.nanolett.2c04220},
    issn = {1530-6984}
}

@article{Timmerman2020CarrierGaN,
    title = {{Carrier dynamics and excitation of {\$}{\textbackslash}mathrm{\{}E{\}}{\{}{\textbackslash}mathrm{\{}u{\}}{\}}{\^{}}{\{}3+{\}}{\$} ions in GaN}},
    year = {2020},
    journal = {Physical Review B},
    author = {Timmerman, Dolf and Mitchell, Brandon and Ichikawa, Shuhei and Nagai, Masaya and Ashida, Masaaki and Fujiwara, Yasufumi},
    number = {24},
    month = {6},
    pages = {245306},
    volume = {101},
    publisher = {American Physical Society},
    url = {https://link.aps.org/doi/10.1103/PhysRevB.101.245306},
    doi = {10.1103/PhysRevB.101.245306}
}

@article{Jobez2015CoherentMemory,
    title = {{Coherent Spin Control at the Quantum Level in an Ensemble-Based Optical Memory}},
    year = {2015},
    journal = {Physical Review Letters},
    author = {Jobez, Pierre and Laplane, Cyril and Timoney, Nuala and Gisin, Nicolas and Ferrier, Alban and Goldner, Philippe and Afzelius, Mikael},
    number = {23},
    month = {6},
    pages = {230502},
    volume = {114},
    publisher = {American Physical Society},
    url = {https://link.aps.org/doi/10.1103/PhysRevLett.114.230502},
    doi = {10.1103/PhysRevLett.114.230502},
    issn = {0031-9007}
}

@article{Mitchell2019Color-TunablilityInjection,
    title = {{Color-Tunablility in GaN LEDs Based on Atomic Emission Manipulation under Current Injection}},
    year = {2019},
    journal = {ACS Photonics},
    author = {Mitchell, Brandon and Wei, Ruoqiao and Takatsu, Junichi and Timmerman, Dolf and Gregorkiewicz, Tom and Zhu, Wanxin and Ichikawa, Shuhei and Tatebayashi, Jun and Fujiwara, Yasufumi and Dierolf, Volkmar},
    number = {5},
    month = {5},
    pages = {1153--1161},
    volume = {6},
    publisher = {American Chemical Society},
    url = {https://pubs.acs.org/sharingguidelines},
    doi = {10.1021/acsphotonics.8b01461},
    issn = {23304022}
}

@article{Vinh2007ConcentrationNanolayers,
    title = {{Concentration of {\$}{\{}{\textbackslash}mathrm{\{}Er{\}}{\}}{\^{}}{\{}3+{\}}{\$} ions contributing to {\$}1.5{\textbackslash}text{\{}{\textbackslash}ensuremath{\{}-{\}}{\}}{\textbackslash}ensuremath{\{}{\textbackslash}mu{\}}{\textbackslash}mathrm{\{}m{\}}{\$} emission in {\$}{\textbackslash}mathrm{\{}Si{\}}∕{\textbackslash}mathrm{\{}Si{\}}:{\textbackslash}mathrm{\{}Er{\}}{\$} nanolayers}},
    year = {2007},
    journal = {Physical Review B},
    author = {Vinh, N Q and Minissale, S and Vrielinck, H and Gregorkiewicz, T},
    number = {8},
    month = {8},
    pages = {85339},
    volume = {76},
    publisher = {American Physical Society},
    url = {https://link.aps.org/doi/10.1103/PhysRevB.76.085339},
    doi = {10.1103/PhysRevB.76.085339}
}

@article{Altahtamouni2015DramaticStructures,
    title = {{Dramatic enhancement of 1.54 {$\mu$}m emission in Er doped GaN quantum well structures}},
    year = {2015},
    journal = {Applied Physics Letters},
    author = {Al tahtamouni, T M and Stachowicz, M and Li, J and Lin, J Y and Jiang, H X},
    number = {12},
    month = {3},
    pages = {121106},
    volume = {106},
    url = {https://doi.org/10.1063/1.4916393},
    doi = {10.1063/1.4916393},
    issn = {0003-6951}
}

@article{Zhong2019EmergingNanophotonics,
    title = {{Emerging rare-earth doped material platforms for quantum nanophotonics}},
    year = {2019},
    journal = {Nanophotonics},
    author = {Zhong, Tian and Goldner, Philippe},
    number = {11},
    month = {11},
    pages = {2003--2015},
    volume = {8},
    publisher = {Walter de Gruyter GmbH},
    url = {https://www.degruyter.com/document/doi/10.1515/nanoph-2019-0185/html},
    doi = {10.1515/NANOPH-2019-0185/ASSET/GRAPHIC/J{\_}NANOPH-2019-0185{\_}FIG{\_}002.JPG},
    issn = {21928614}
}

@article{Iwaya2023EnhancedAnnealing,
    title = {{Enhanced light output of Eu, O-codoped GaN caused by reconfiguration of luminescent sites during post-growth thermal annealing}},
    year = {2023},
    journal = {Applied Physics Letters},
    author = {Iwaya, T and Ichikawa, S and Timmerman, D and Tatebayashi, J and Fujiwara, Y},
    number = {3},
    month = {1},
    pages = {32102},
    volume = {122},
    url = {https://doi.org/10.1063/5.0136880},
    doi = {10.1063/5.0136880},
    issn = {0003-6951}
}

@article{Murakami2023EnhancedSpectroscopy,
    title = {{Enhanced luminescence efficiency in Eu-doped GaN superlattice structures revealed by terahertz emission spectroscopy}},
    year = {2023},
    journal = {Communications Materials},
    author = {Murakami, Fumikazu and Takeo, Atsushi and Mitchell, Brandon and Dierolf, Volkmar and Fujiwara, Yasufumi and Tonouchi, Masayoshi},
    number = {1},
    pages = {100},
    volume = {4},
    url = {https://doi.org/10.1038/s43246-023-00428-6},
    doi = {10.1038/s43246-023-00428-6},
    issn = {2662-4443}
}

@article{Zhu2016EnhancedEnvironment,
    title = {{Enhanced photo/electroluminescence properties of Eu-doped GaN through optimization of the growth temperature and Eu related defect environment}},
    year = {2016},
    journal = {APL Materials},
    author = {Zhu, W and Mitchell, B and Timmerman, D and Uedono, A and Koizumi, A and Fujiwara, Y},
    number = {5},
    month = {5},
    pages = {56103},
    volume = {4},
    url = {https://doi.org/10.1063/1.4950826},
    doi = {10.1063/1.4950826},
    issn = {2166-532X}
}

@article{Ichikawa2021EnhancedHoles,
    title = {{Enhanced Red Emission of Eu,O-Codoped Ga N Embedded in a Photonic Crystal Nanocavity with Hexagonal Air Holes}},
    year = {2021},
    journal = {Physical Review Applied},
    author = {Ichikawa, Shuhei and Sasaki, Yutaka and Iwaya, Takenori and Murakami, Masato and Ashida, Masaaki and Timmerman, Dolf and Tatebayashi, Jun and Fujiwara, Yasufumi},
    number = {3},
    month = {3},
    volume = {15},
    publisher = {American Physical Society},
    doi = {10.1103/PhysRevApplied.15.034086},
    issn = {23317019}
}

@article{Ichikawa2021Eu-dopedGamut,
    title = {{Eu-doped GaN and InGaN monolithically stacked full-color LEDs with a wide color gamut}},
    year = {2021},
    journal = {Applied Physics Express},
    author = {Ichikawa, Shuhei and Shiomi, Keishi and Morikawa, Takaya and Timmerman, Dolf and Sasaki, Yutaka and Tatebayashi, Jun and Fujiwara, Yasufumi},
    number = {3},
    month = {3},
    pages = {031008},
    volume = {14},
    url = {https://iopscience.iop.org/article/10.35848/1882-0786/abe603},
    doi = {10.35848/1882-0786/abe603},
    issn = {1882-0778}
}

@article{Woodward2011ExcitationCenter,
    title = {{Excitation of Eu3+ in gallium nitride epitaxial layers: Majority versus trap defect center}},
    year = {2011},
    journal = {Applied Physics Letters},
    author = {Woodward, N. and Poplawsky, J. and Mitchell, B. and Nishikawa, A. and Fujiwara, Y. and Dierolf, V.},
    number = {1},
    pages = {3--5},
    volume = {98},
    doi = {10.1063/1.3533806},
    issn = {00036951}
}

@article{Fleischman2009ExcitationSpectroscopy,
    title = {{Excitation pathways and efficiency of Eu ions in GaN by site-selective spectroscopy}},
    year = {2009},
    journal = {Applied Physics B},
    author = {Fleischman, Z and Munasinghe, C and Steckl, A J and Wakahara, A and Zavada, J and Dierolf, V},
    number = {3},
    pages = {607--618},
    volume = {97},
    url = {https://doi.org/10.1007/s00340-009-3605-x},
    doi = {10.1007/s00340-009-3605-x},
    issn = {1432-0649}
}

@article{Reshchikov2020GiantGaN,
    title = {{Giant shifts of photoluminescence bands in GaN}},
    year = {2020},
    journal = {Journal of Applied Physics},
    author = {Reshchikov, Michael A.},
    number = {5},
    month = {2},
    volume = {127},
    publisher = {American Institute of Physics Inc.},
    doi = {10.1063/1.5140686},
    issn = {10897550}
}

@article{Zhu2017High-PowerEpitaxy,
    title = {{High-Power Eu-Doped GaN Red LED Based on a Multilayer Structure Grown at Lower Temperatures by Organometallic Vapor Phase Epitaxy}},
    year = {2017},
    journal = {MRS Advances},
    author = {Zhu, W and Mitchell, B and Timmerman, D and Koizumi, A and Gregorkiewicz, T and Fujiwara, Y},
    edition = {2017/01/19},
    number = {3},
    pages = {159--164},
    volume = {2},
    publisher = {Materials Research Society},
    url = {https://www.cambridge.org/core/product/D49D619180721161FD50489C5C91F9F8},
    doi = {DOI: 10.1557/adv.2017.67}
}

@article{Iwaya2022ImprovedEngineering,
    title = {{Improved Q-factors of III-nitride-based photonic crystal nanocavities by optical loss engineering}},
    year = {2022},
    journal = {Optics Express},
    author = {Iwaya, Takenori and Ichikawa, Shuhei and Timmerman, Dolf and Tatebayashi, Jun and Fujiwara, Yasufumi},
    number = {16},
    pages = {28853--28864},
    volume = {30},
    publisher = {Optica Publishing Group},
    url = {https://opg.optica.org/oe/abstract.cfm?URI=oe-30-16-28853},
    doi = {10.1364/OE.460467}
}

@article{Ourari2023IndistinguishableState,
    title = {{Indistinguishable telecom band photons from a single Er ion in the solid state}},
    year = {2023},
    journal = {Nature},
    author = {Ourari, Salim and Dusanowski, {\L}ukasz and Horvath, Sebastian P and Uysal, Mehmet T and Phenicie, Christopher M and Stevenson, Paul and Raha, Mouktik and Chen, Songtao and Cava, Robert J and de Leon, Nathalie P and Thompson, Jeff D},
    number = {7976},
    pages = {977--981},
    volume = {620},
    url = {https://doi.org/10.1038/s41586-023-06281-4},
    doi = {10.1038/s41586-023-06281-4},
    issn = {1476-4687}
}

@article{Nilsson2002InitialCrystals,
    title = {{Initial Experiments Concerning Quantum Information Processing in Rare-Earth-Ion Doped Crystals}},
    year = {2002},
    journal = {Physica Scripta},
    author = {Nilsson, M. and Rippe, L. and Ohlsson, N. and Christiansson, T. and Kr{\"{o}}ll, S.},
    number = {1},
    pages = {178},
    volume = {T102},
    url = {https://iopscience.iop.org/article/10.1238/Physica.Topical.102a00178},
    doi = {10.1238/Physica.Topical.102a00178},
    issn = {0031-8949}
}

@article{Binnemans2015InterpretationSpectra,
    title = {{Interpretation of europium(III) spectra}},
    year = {2015},
    journal = {Coordination Chemistry Reviews},
    author = {Binnemans, Koen},
    month = {7},
    pages = {1--45},
    volume = {295},
    publisher = {Elsevier},
    doi = {10.1016/J.CCR.2015.02.015},
    issn = {0010-8545}
}

@article{Wakamatsu2013LuminescenceSites,
    title = {{Luminescence properties of Eu-doped GaN under resonant excitation and quantitative evaluation of luminescent sites}},
    year = {2013},
    journal = {Journal of Applied Physics},
    author = {Wakamatsu, Ryuta and Lee, Dong Gun and Koizumi, Atsushi and Dierolf, Volkmar and Fujiwara, Yasufumi},
    number = {4},
    pages = {5--9},
    volume = {114},
    url = {http://dx.doi.org/10.1063/1.4816088},
    doi = {10.1063/1.4816088},
    issn = {00218979}
}

@article{Vinh2003MicroscopicSilicon,
    title = {{Microscopic Structure of Er-Related Optically Active Centers in Crystalline Silicon}},
    year = {2003},
    journal = {Physical Review Letters},
    author = {Vinh, N Q and Przybyli{\'{n}}ska, H and Krasil’nik, Z F and Gregorkiewicz, T},
    number = {6},
    month = {2},
    pages = {66401},
    volume = {90},
    publisher = {American Physical Society},
    url = {https://link.aps.org/doi/10.1103/PhysRevLett.90.066401},
    doi = {10.1103/PhysRevLett.90.066401}
}

@article{Austin2022ModelingLayers,
    title = {{Modeling defect mediated color-tunability in LEDs with Eu-doped GaN-based active layers}},
    year = {2022},
    journal = {Journal of Applied Physics},
    author = {Austin, Hayley J and Mitchell, Brandon and Timmerman, Dolf and Tatebayashi, Jun and Ichikawa, Shuhei and Fujiwara, Yasufumi and Dierolf, Volkmar},
    number = {4},
    month = {1},
    pages = {045701},
    volume = {131},
    url = {https://doi.org/10.1063/5.0077223},
    doi = {10.1063/5.0077223},
    issn = {0021-8979}
}

@article{Ma2023MonteCrystals,
    title = {{Monte Carlo simulation of the nuclear spin decoherence process in {\$}{\{}{\textbackslash}mathrm{\{}Eu{\}}{\}}{\^{}}{\{}3+{\}}{\$}:{\$}{\{}{\textbackslash}mathrm{\{}Y{\}}{\}}{\_}{\{}2{\}}{\{}{\textbackslash}mathrm{\{}SiO{\}}{\}}{\_}{\{}5{\}}{\$} crystals}},
    year = {2023},
    journal = {Physical Review B},
    author = {Ma, You-Zhi and Lv, You-Cai and Yang, Tian-Shu and Ma, Yu and Zhou, Zong-Quan and Li, Chuan-Feng and Guo, Guang-Can},
    number = {1},
    month = {1},
    pages = {14310},
    volume = {107},
    publisher = {American Physical Society},
    url = {https://link.aps.org/doi/10.1103/PhysRevB.107.014310},
    doi = {10.1103/PhysRevB.107.014310}
}

@article{Sekiguchi2019ObservationEpitaxy,
    title = {{Observation of single optical site of Eu and Mg codoped GaN grown by NH3-source molecular beam epitaxy}},
    year = {2019},
    journal = {Journal of Applied Physics},
    author = {Sekiguchi, Hiroto and Sakai, Masaru and Kamada, Takuho and Yamane, Keisuke and Okada, Hiroshi and Wakahara, Akihiro},
    number = {17},
    month = {5},
    pages = {175702},
    volume = {125},
    url = {https://doi.org/10.1063/1.5090893},
    doi = {10.1063/1.5090893},
    issn = {0021-8979}
}

@article{Ma2021One-hourMemory,
    title = {{One-hour coherent optical storage in an atomic frequency comb memory}},
    year = {2021},
    journal = {Nature Communications},
    author = {Ma, Yu and Ma, You-Zhi and Zhou, Zong-Quan and Li, Chuan-Feng and Guo, Guang-Can},
    number = {1},
    pages = {2381},
    volume = {12},
    url = {https://doi.org/10.1038/s41467-021-22706-y},
    doi = {10.1038/s41467-021-22706-y},
    issn = {2041-1723}
}

@article{Vinh2004OpticalNanostructures,
    title = {{Optical properties of a single type of optically active center in {\$}{\textbackslash}mathrm{\{}Si{\}}∕{\textbackslash}mathrm{\{}Si{\}}:{\textbackslash}mathrm{\{}Er{\}}{\$} nanostructures}},
    year = {2004},
    journal = {Physical Review B},
    author = {Vinh, N Q and Przybyli{\'{n}}ska, H and Krasil’nik, Z F and Gregorkiewicz, T},
    number = {11},
    month = {9},
    pages = {115332},
    volume = {70},
    publisher = {American Physical Society},
    url = {https://link.aps.org/doi/10.1103/PhysRevB.70.115332},
    doi = {10.1103/PhysRevB.70.115332}
}

@article{Zhong2015OpticallyTime,
    title = {{Optically addressable nuclear spins in a solid with a six-hour coherence time}},
    year = {2015},
    journal = {Nature},
    author = {Zhong, Manjin and Hedges, Morgan P and Ahlefeldt, Rose L and Bartholomew, John G and Beavan, Sarah E and Wittig, Sven M and Longdell, Jevon J and Sellars, Matthew J},
    number = {7533},
    pages = {177--180},
    volume = {517},
    url = {https://doi.org/10.1038/nature14025},
    doi = {10.1038/nature14025},
    issn = {1476-4687}
}

@article{Mitchell2018Perspective:Doping,
    title = {{Perspective: Toward efficient GaN-based red light emitting diodes using europium doping}},
    year = {2018},
    journal = {Journal of Applied Physics},
    author = {Mitchell, Brandon and Dierolf, Volkmar and Gregorkiewicz, Tom and Fujiwara, Yasufumi},
    number = {16},
    month = {4},
    pages = {1--12},
    volume = {123},
    url = {https://pubs.aip.org/jap/article/123/16/160901/400705/Perspective-Toward-efficient-GaN-based-red-light},
    doi = {10.1063/1.5010762},
    issn = {0021-8979}
}

@article{Katchkanov2005PhotoluminescenceEpilayers,
    title = {{Photoluminescence studies of Eu-implanted GaN epilayers}},
    year = {2005},
    journal = {physica status solidi (b)},
    author = {Katchkanov, V and O'Donnell, K P and Dalmasso, S and Martin, R W and Braud, A and Nakanishi, Y and Wakahara, A and Yoshida, A},
    number = {7},
    month = {6},
    pages = {1491--1496},
    volume = {242},
    publisher = {John Wiley {\&} Sons, Ltd},
    url = {https://doi.org/10.1002/pssb.200440032},
    doi = {https://doi.org/10.1002/pssb.200440032},
    issn = {0370-1972}
}

@article{Wei2019PicosecondIons,
    title = {{Picosecond time-resolved dynamics of energy transfer between GaN and the various excited states of {\$}{\textbackslash}mathrm{\{}E{\}}{\{}{\textbackslash}mathrm{\{}u{\}}{\}}{\^{}}{\{}3+{\}}{\$} ions}},
    year = {2019},
    journal = {Physical Review B},
    author = {Wei, Ruoqiao and Mitchell, Brandon and Timmerman, Dolf and Gregorkiewicz, Tom and Zhu, Wanxin and Tatebayashi, Jun and Ichikawa, Shuhei and Fujiwara, Yasufumi and Dierolf, Volkmar},
    number = {8},
    month = {8},
    pages = {81201},
    volume = {100},
    publisher = {American Physical Society},
    url = {https://link.aps.org/doi/10.1103/PhysRevB.100.081201},
    doi = {10.1103/PhysRevB.100.081201}
}

@article{Stepikhova2001Properties320323,
    title = {{Properties of optically active Si:Er and Si1−xGex layers grown by the sublimation MBE method: [Thin Solid Films 369 (2000) 320–323],}},
    year = {2001},
    journal = {Thin Solid Films},
    author = {Stepikhova, M. V. and Andreev, B. A. and Shmagin, V. B. and Krasil'Nik, Z. F. and Kuznetsov, V. P. and Shengurov, V. G. and Svetlov, S. P. and Jantsch, W. and Palmetshofer, L. and Ellmer, H.},
    number = {1},
    month = {1},
    pages = {164--169},
    volume = {381},
    publisher = {Elsevier},
    doi = {10.1016/S0040-6090(00)01373-0},
    issn = {0040-6090}
}

@article{Guo2023Rare-earthQuo,
    title = {{Rare-earth quantum memories: The experimental status quo}},
    year = {2023},
    journal = {Frontiers of Physics},
    author = {Guo, Mucheng and Liu, Shuping and Sun, Weiye and Ren, Miaomiao and Wang, Fudong and Zhong, Manjin},
    number = {2},
    pages = {21303},
    volume = {18},
    url = {https://doi.org/10.1007/s11467-022-1240-8},
    doi = {10.1007/s11467-022-1240-8},
    issn = {2095-0470}
}

@article{Thiel2011Rare-earth-dopedProcessing,
    title = {{Rare-earth-doped materials for applications in quantum information storage and signal processing}},
    year = {2011},
    journal = {Journal of Luminescence},
    author = {Thiel, C W and B{\"{o}}ttger, Thomas and Cone, R L},
    number = {3},
    pages = {353--361},
    volume = {131},
    url = {https://www.sciencedirect.com/science/article/pii/S002223131000534X},
    doi = {https://doi.org/10.1016/j.jlumin.2010.12.015},
    issn = {0022-2313}
}

@article{Zhu2018Re-ExcitationExcitation,
    title = {{Re-Excitation of Trivalent Europium Ions Doped into Gallium Nitride Revealed through Photoluminescence under Pulsed Laser Excitation}},
    year = {2018},
    journal = {ACS Photonics},
    author = {Zhu, Wanxin and Wei, Ruoqiao and Timmerman, Dolf and Gregorkiewicz, Tom and Mitchell, Brandon and Fujiwara, Yasufumi and Dierolf, Volkmar},
    number = {3},
    month = {3},
    pages = {875--880},
    volume = {5},
    publisher = {American Chemical Society},
    url = {https://doi.org/10.1021/acsphotonics.7b01090},
    doi = {10.1021/acsphotonics.7b01090}
}

@article{Ho2018Room-TemperatureRegion,
    title = {{Room-Temperature Lasing Action in GaN Quantum Wells in the Infrared 1.5 {$\mu$}m Region}},
    year = {2018},
    journal = {ACS Photonics},
    author = {Ho, V X and Al tahtamouni, T M and Jiang, H X and Lin, J Y and Zavada, J M and Vinh, N Q},
    number = {4},
    month = {4},
    pages = {1303--1309},
    volume = {5},
    publisher = {American Chemical Society},
    url = {https://doi.org/10.1021/acsphotonics.7b01253},
    doi = {10.1021/acsphotonics.7b01253}
}

@article{Woodward2011SiteLayers,
    title = {{Site and sample dependent electron-phonon coupling of Eu ions in epitaxial-grown GaN layers}},
    year = {2011},
    journal = {Optical Materials},
    author = {Woodward, N. and Nishikawa, A. and Fujiwara, Y. and Dierolf, V.},
    number = {7},
    pages = {1050--1054},
    volume = {33},
    publisher = {Elsevier B.V.},
    url = {http://dx.doi.org/10.1016/j.optmat.2010.09.029},
    doi = {10.1016/j.optmat.2010.09.029},
    issn = {09253467}
}

@article{Kuznetsov2000SpecialStructures,
    title = {{Special features of the sublimational molecular-beam epitaxy of Si and its potentialities for growing Si:Er/Si structures}},
    year = {2000},
    journal = {Semiconductors},
    author = {Kuznetsov, V P and Rubtsova, R A},
    number = {5},
    pages = {502--509},
    volume = {34},
    url = {https://doi.org/10.1134/1.1188015},
    doi = {10.1134/1.1188015},
    issn = {1090-6479}
}

@article{Arcangeli2014Spectroscopy5,
    title = {{Spectroscopy and coherence lifetime extension of hyperfine transitions in    151  Eu    3 +    :Y 2   SiO 5}},
    year = {2014},
    journal = {Physical Review B},
    author = {Arcangeli, Andrea and Lovri{\'{c}}, Marko and Tumino, Biagio and Ferrier, Alban and Goldner, Philippe},
    number = {18},
    month = {5},
    pages = {184305},
    volume = {89},
    publisher = {American Physical Society},
    url = {https://link.aps.org/doi/10.1103/PhysRevB.89.184305},
    doi = {10.1103/PhysRevB.89.184305},
    issn = {1098-0121}
}

@article{Uysal2025Spin-PhotonBand,
    title = {{Spin-Photon Entanglement of a Single {\$}{\{}{\textbackslash}mathrm{\{}Er{\}}{\}}{\^{}}{\{}3+{\}}{\$} Ion in the Telecom Band}},
    year = {2025},
    journal = {Physical Review X},
    author = {Uysal, Mehmet T and Dusanowski, Lukasz and Xu, Haitong and Horvath, Sebastian P and Ourari, Salim and Cava, Robert J and de Leon, Nathalie P and Thompson, Jeff D},
    number = {1},
    month = {3},
    pages = {11071},
    volume = {15},
    publisher = {American Physical Society},
    url = {https://link.aps.org/doi/10.1103/PhysRevX.15.011071},
    doi = {10.1103/PhysRevX.15.011071}
}

@inproceedings{Copelman2020StrongGaN,
    title = {{Strong crystal field splitting and polarization dependence observed in the emission from Eu3+ ions doped into GaN}},
    year = {2020},
    booktitle = {Proc.SPIE},
    author = {Copelman, S and Austin, H and Timmerman, D and Poplawsky, J D and Waite, M and Tatebayashi, J and Ichikawa, S and Fujiwara, Y and Dierolf, V and Mitchell, B},
    month = {2},
    pages = {113021Z},
    volume = {11302},
    url = {https://doi.org/10.1117/12.2544005},
    doi = {10.1117/12.2544005}
}

@article{Mitchell2017SynthesisEpitaxy,
    title = {{Synthesis and characterization of a liquid Eu precursor (EuCppm2) allowing for valence control of Eu ions doped into GaN by organometallic vapor phase epitaxy}},
    year = {2017},
    journal = {Materials Chemistry and Physics},
    author = {Mitchell, Brandon and Koizumi, Atsushi and Nunokawa, Takumi and Wakamatsu, Ryuta and Lee, Dong-gun and Saitoh, Yasuhisa and Timmerman, Dolf and Kuboshima, Yoshinori and Mogi, Takayuki and Higashi, Shintaro and Kikukawa, Kaoru and Ofuchi, Hironori and Honma, Tetsuo and Fujiwara, Yasufumi},
    pages = {140--146},
    volume = {193},
    url = {https://www.sciencedirect.com/science/article/pii/S0254058417301529},
    doi = {https://doi.org/10.1016/j.matchemphys.2017.02.021},
    issn = {0254-0584}
}

@article{Konz2003Temperature5,
    title = {{Temperature and concentration dependence of optical dephasing, spectral-hole lifetime, and anisotropic absorption in     Eu   3 +       : Y   2       SiO   5}},
    year = {2003},
    journal = {Physical Review B},
    author = {K{\"{o}}nz, Flurin and Sun, Y. and Thiel, C. W. and Cone, R. L. and Equall, R. W. and Hutcheson, R. L. and Macfarlane, R. M.},
    number = {8},
    month = {8},
    pages = {085109},
    volume = {68},
    publisher = {American Physical Society},
    url = {https://link.aps.org/doi/10.1103/PhysRevB.68.085109},
    doi = {10.1103/PhysRevB.68.085109},
    issn = {0163-1829}
}

@article{Reshchikov2018ThermalGaN,
    title = {{Thermal quenching of the yellow luminescence in GaN}},
    year = {2018},
    journal = {Journal of Applied Physics},
    author = {Reshchikov, M. A. and Albarakati, N. M. and Monavarian, M. and Avrutin, V. and Morko{\c{c}}, H.},
    number = {16},
    month = {4},
    volume = {123},
    publisher = {American Institute of Physics Inc.},
    doi = {10.1063/1.4995275},
    issn = {10897550}
}

@article{Serrano2022Ultra-narrowCrystals,
    title = {{Ultra-narrow optical linewidths in rare-earth molecular crystals}},
    year = {2022},
    journal = {Nature},
    author = {Serrano, Diana and Kuppusamy, Senthil Kumar and Heinrich, Benoît and Fuhr, Olaf and Hunger, David and Ruben, Mario and Goldner, Philippe},
    number = {7900},
    pages = {241--246},
    volume = {603},
    url = {https://doi.org/10.1038/s41586-021-04316-2},
    doi = {10.1038/s41586-021-04316-2},
    issn = {1476-4687}
}

@article{Yu2022Ultra-smallLithography,
    title = {{Ultra-small size (1–20 {$\mu$}m) blue and green micro-LEDs fabricated by laser direct writing lithography}},
    year = {2022},
    journal = {Applied Physics Letters},
    author = {Yu, Luming and Lu, Boyang and Yu, Ping and Wang, Yang and Ding, Guojian and Feng, Qi and Jiang, Yang and Chen, Hong and Huang, Kai and Hao, Zhibiao and Yu, Jiadong and Luo, Yi and Sun, Changzheng and Xiong, Bing and Han, Yanjun and Wang, Jian and Li, Hongtao and Wang, Lai},
    number = {4},
    month = {7},
    pages = {42106},
    volume = {121},
    url = {https://doi.org/10.1063/5.0099642},
    doi = {10.1063/5.0099642},
    issn = {0003-6951}
}

@article{Yano1991UltralongEu3+:Y2SiO5,
    title = {{Ultralong optical dephasing time in Eu3+:Y2SiO5}},
    year = {1991},
    journal = {Optics Letters},
    author = {Yano, Ryuzi and Mitsunaga, Masaharu and Uesugi, Naoshi},
    number = {23},
    pages = {1884--1886},
    volume = {16},
    publisher = {Optica Publishing Group},
    url = {https://opg.optica.org/ol/abstract.cfm?URI=ol-16-23-1884},
    doi = {10.1364/OL.16.001884}
}

@article{Mitchell2016UtilizationApplications,
    title = {{Utilization of native oxygen in Eu(RE)-doped GaN for enabling device compatibility in optoelectronic applications}},
    year = {2016},
    journal = {Scientific Reports},
    author = {Mitchell, B and Timmerman, D and Poplawsky, J and Zhu, W and Lee, D and Wakamatsu, R and Takatsu, J and Matsuda, M and Guo, W and Lorenz, K and Alves, E and Koizumi, A and Dierolf, V and Fujiwara, Y},
    number = {1},
    pages = {18808},
    volume = {6},
    url = {https://doi.org/10.1038/srep18808},
    doi = {10.1038/srep18808},
    issn = {2045-2322}
}

\end{document}


\title{Supplementary Information:\\Site-selective enhancement of Eu emission in delta-doped GaN}

\author{Amelia R. Klein}
\affiliation{Quantum Engineering Laboratory, Department of Electrical and Systems Engineering, University of Pennsylvania, 200 S. 33rd St. Philadelphia, Pennsylvania, 19104, USA}

\author{Hayley J. Austin}
\affiliation{Department of Physics, Lehigh University, Bethlehem, Pennsylvania 18015, USA}

\author{Fumikazu Murakami}
\affiliation{Institute of Laser Engineering, Osaka University, 2-6 Yamada-oka, Suita, Osaka 565-0871, Japan}

\author{Jamie Ford}
\affiliation{Singh Center for Nanotechnology, University of Pennsylvania, Philadelphia, Pennsylvania, 19104, USA}

\author{Jun Tatebayashi}
\affiliation{Department of Materials and Manufacturing Science, Graduate School of Engineering, Osaka University, 2-1 Yamada-oka, Suita, Osaka 565-0871, Japan}

\author{Masayoshi Tonouchi}
\affiliation{Institute of Laser Engineering, Osaka University, 2-6 Yamada-oka, Suita, Osaka 565-0871, Japan}

\author{Yasufumi Fujiwara}
\affiliation{Intra-Photonics Research Center, Research Organization of Science and Technology, Ritsumeikan University, 1-1-1 Nojihigashi, Kusatsu, Shiga 525-8577, Japan}
\affiliation{Department of Materials and Manufacturing Science, Graduate School of Engineering, Osaka University, 2-1 Yamada-oka, Suita, Osaka 565-0871, Japan}

\author{Volkmar Dierolf}
\affiliation{Department of Physics, Lehigh University, Bethlehem, Pennsylvania 18015, USA}

\author{Lee C. Bassett}
\affiliation{Quantum Engineering Laboratory, Department of Electrical and Systems Engineering, University of Pennsylvania, 200 S. 33rd St. Philadelphia, Pennsylvania, 19104, USA}

\author{Brandon Mitchell}
\affiliation{Department of Physics, West Chester University, West Chester, Pennsylvania 19383, USA}
\affiliation{Department of Physics, Lehigh University, Bethlehem, Pennsylvania 18015, USA}
\affiliation{Department of Materials and Manufacturing Science, Graduate School of Engineering, Osaka University, 2-1 Yamada-oka, Suita, Osaka 565-0871, Japan}
\email{bmitchell@wcupa.edu}

\date{\today}

\maketitle



\newpage

\begin{table}
\centering
\begin{tabular}{|c|c|c|c|c|c|c|c|}
\hline
Sample & substrate & temperature (\textdegree C) & $\mathrm{NH_3}$ (SLM) & TMG (sccm) & V/III ratio & Eu (SLM) & Active layer growth time\\
\hline
UD & c-sapphire & 960 & 4.0 & 5.25 & 6985 & \cellcolor{yellow}0.7 & \cellcolor{yellow}GaN:Eu = 21 min\\
\hline
10:10 DD & c-sapphire & 960 & 4.0 & 5.25 & 6985 & \cellcolor{yellow}1.5 & \cellcolor{yellow}GaN:Eu/GaN = 38s/40s, x40\\
\hline
10:2 DD & c-sapphire & 960 & 4.0 & 5.25 & 6985 & \cellcolor{yellow}1.5 & \cellcolor{yellow}GaN:Eu/GaN = 7s/40s, x40\\
\hline
10:1 DD & c-sapphire & 960 & 4.0 & 5.25 & 6985 & \cellcolor{yellow}1.3 & \cellcolor{yellow}GaN:Eu/GaN = 4s/40s, x40\\
\hline
\end{tabular}
\label{table:Growth}
\caption{Growth parameters and relevant flow rates used to create the samples studied in this work using organometallic vapor-phase epitaxy. Parameters that differ between samples are highlighted in yellow.}
\end{table}

\section{SIMS Measurements}

Time-of-flight secondary ion mass spectrometry (ToF-SIMS) measurements were performed using a FIB-SEM (TESCAN S8000X).
The samples were irradiated over a 50 $\mu$m square using a 1500 pA $\mathrm{Xe^+}$ plasma beam while overscanned with a 5 kV, 5 nA electron beam.
SEM measurements were taken after the SIMS measurements to confirm the total etched depth, and a consistent etching rate beween 0.5-0.6 nm per measurement frame was observed for all four samples.
ToF-SIMS measurements shown in the main text were taken on a cryogenic stage cooled to -50 C using liquid nitrogen.
Counts of the $m/q$ peak of 151 corresponding to the $\mathrm{~^{151}Eu}$ isotope were used for analysis; the $\mathrm{~^{153}Eu}$ isotope was not used because of overlap with a different signal that became relevant at the low concentrations of the 10:1 DD sample.

\begin{figure}
    \includegraphics{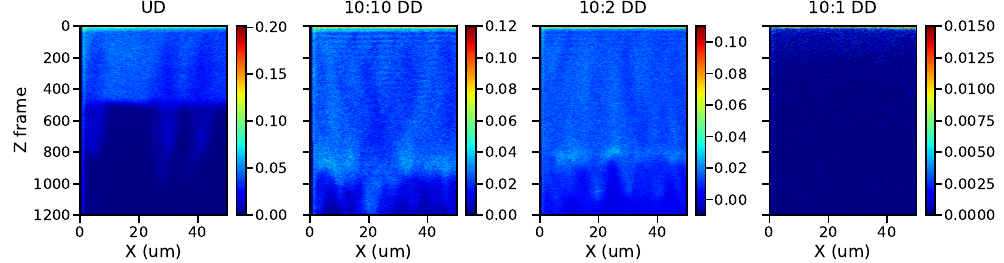}
    \caption{Cross-sectional SIMS data taken at room temperature showing count rates of the detected $\mathrm{~^{151}Eu}$ peak. Clear doped and undoped layers are visible for the 10:10 DD and 10:2 DD samples, especially towards the top of the samples where there are fewer artifacts. The colorbar scales denote the counts per time-of-flight extraction.}
    \label{fig:sims_side_profiles}
\end{figure}

\begin{figure}
    \includegraphics[scale=0.125]{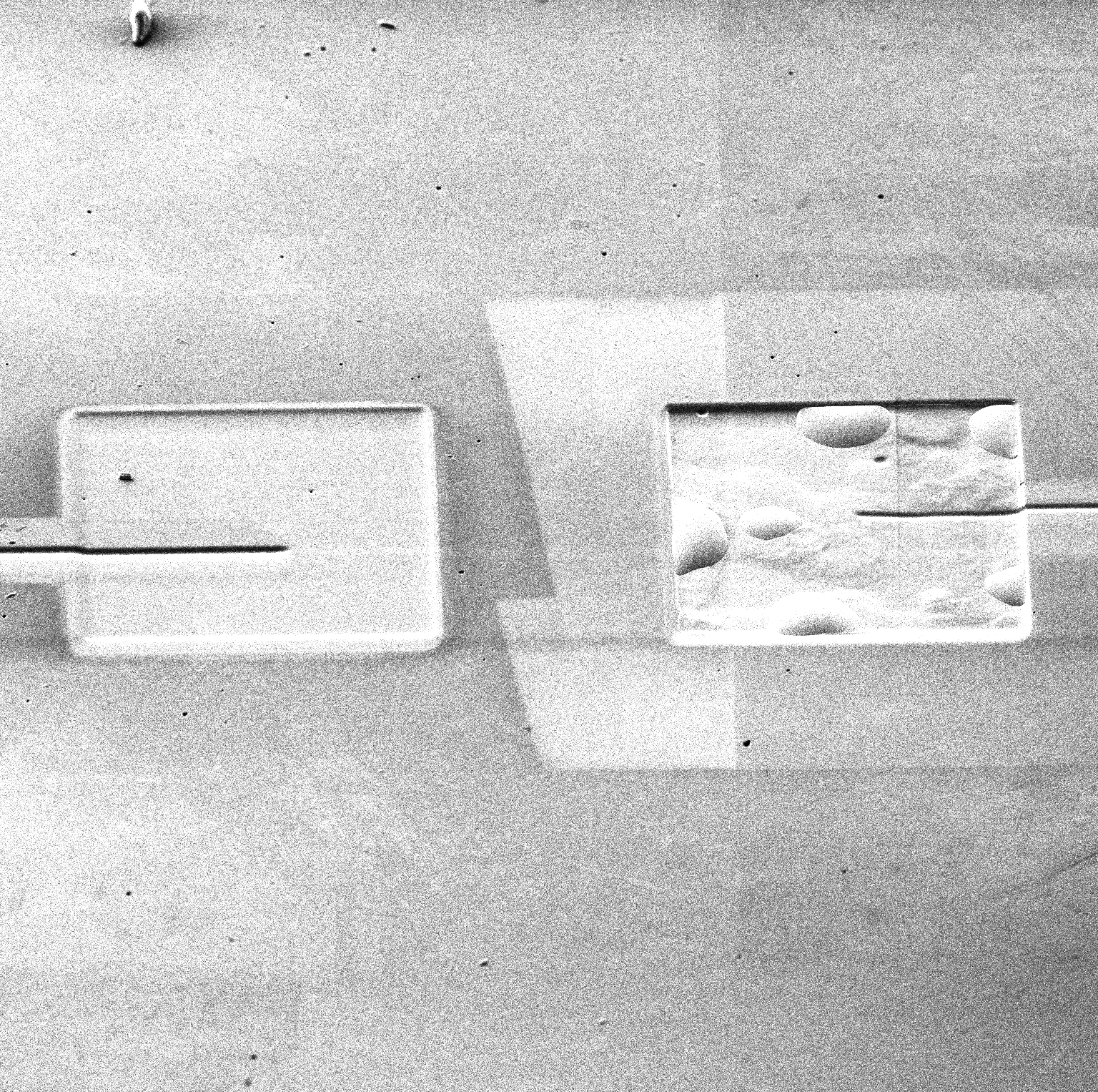}
    \caption{SEM image of 10:1 DD after performing two SIMS measurements. On the left, a measurement performed at -50 \textdegree C resulted in a uniformly etched region. On the right, a measurement performed at room temperature resulted in the formation of numerous bubbles on the surface that appeared during the focused ion beam etching and persisted after the measurement was completed.}
    \label{fig:sims_bubbles_sem}
\end{figure}

Additional SIMS measurements were also taken at room temperature using otherwise the same parameters. 
At room temperature, the overall count rate was higher, allowing for better resolution of the doped and undoped layers.
Cross-sectional profiles of these room-temperature are shown in Fig. \ref{fig:sims_side_profiles}.
While the overall count rates are higher, these room-temperature measurements experienced significant artifacts due to the formation of some sort of ``bubbles'' during the ion beam irradiation.
These bubbles appeared within the first few seconds of the room-temperature SIMS measurements and visibly moved and combined over the course of the measurement in a manner that resembled water droplets.
In SEM images, they appeared as clear lumps that left behind visible trails following the path in which they moved.
In the SIMS images, these bubbles appeared as dark spots in which negligible counts of any element were obtained.
These bubbles persisted over the course of weeks outside of vacuum.
Fig. \ref{fig:sims_bubbles_sem} shows an SEM image in which the resulting etched regions from a room-temperature and low-temperature SIMS measurements were taken on the 10:1 DD sample, showing a clean etched region when measured at low temperatures and several bubbles with trails when measured at room temperature.
Because of the artifacts left by these bubbles, these room-temperature SIMS measurements are unsuitable for the quantitative analysis used in the main text, but are included here as evidence of the existence of doped and undoped layers.

\begin{figure}
    \includegraphics{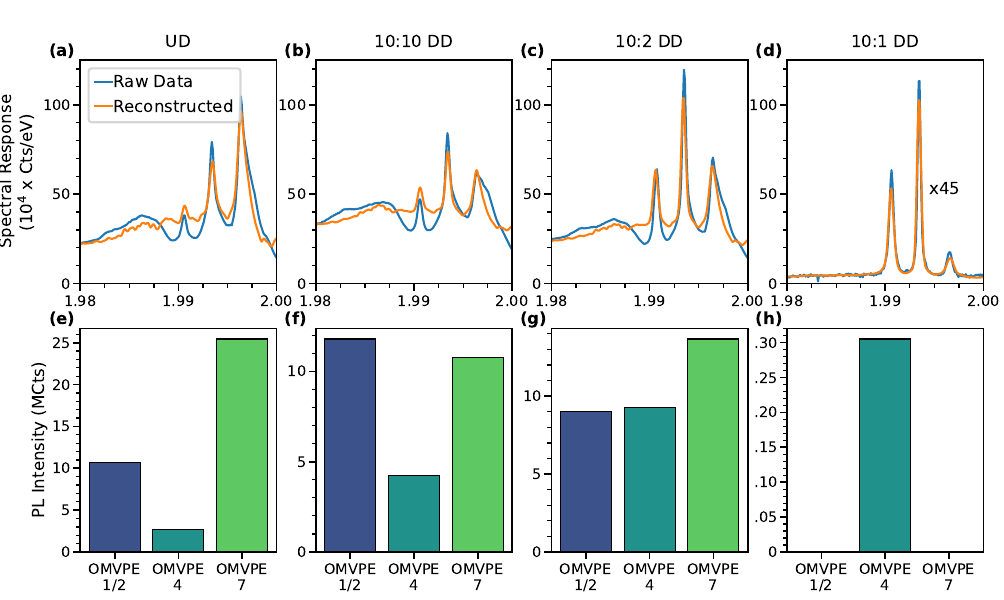}
    \caption{Above bandgap spectral decomposition.
    \textbf{(a)-(d)}: PL spectra at 20K taken with UV excitation above the GaN bandgap at 351 nm (blue curves), fit to the same three spectral components extracted from the CEES maps using spectral decomposition, plus a constant background (orange curves).
    \textbf{(e)-(h)}: Amplitudes of the components used to generate the fits.}
    \label{fig:spectral_decomposition_abg}
\end{figure}

\begin{figure}
    \includegraphics{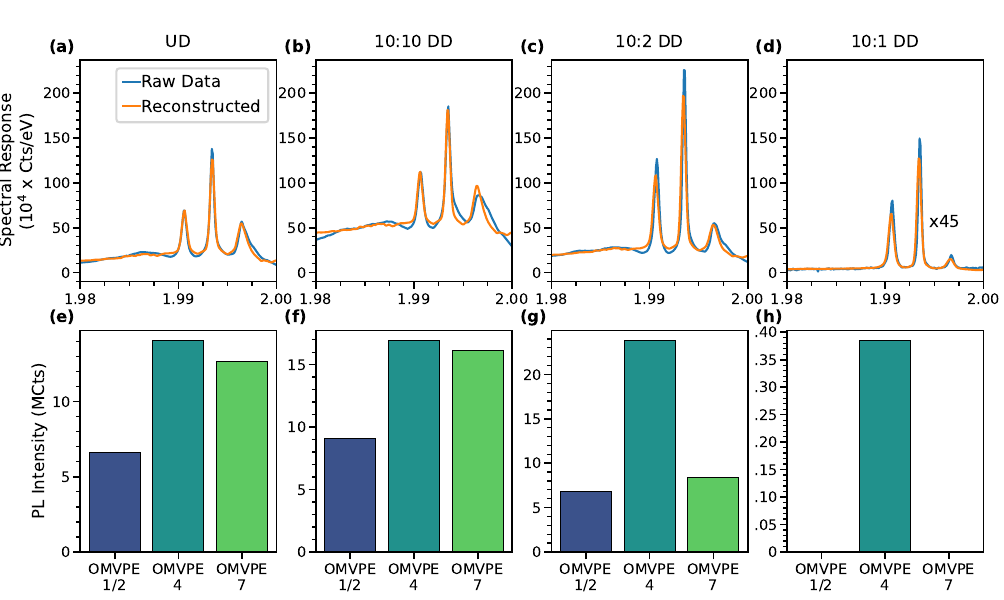}
    \caption{Below bandgap spectral decomposition.
    \textbf{(a)-(d)}: PL spectra at 20K taken with UV excitation just below the GaN bandgap at 364 nm (blue curves), fit to the same three spectral components extracted from the CEES maps using spectral decomposition, plus a constant background (orange curves).
    \textbf{(e)-(h)}: Amplitudes of the components used to generate the fits.}
    \label{fig:spectral_decomposition_bbg}
\end{figure}

\begin{figure}
    \includegraphics{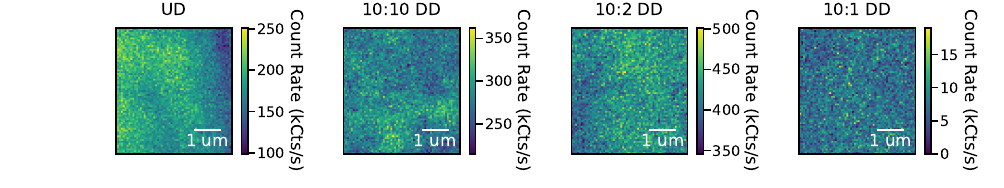}
    \caption{Spatial photoluminescence scans of representative regions of all four samples using excitation resonant with OMVPE4.}
    \label{fig:pl_scans}
\end{figure}

\begin{figure}
    \includegraphics{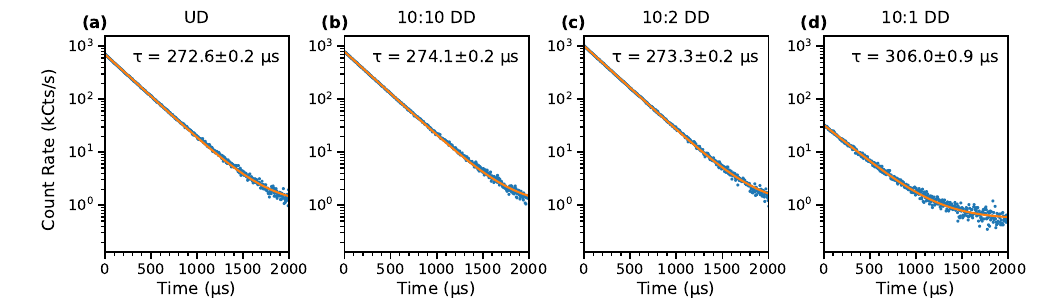}
    \caption{Lifetime measurements of all four samples taken with phonon-assisted resonant excitation of OMVPE4 (571.18 nm) at 4K. Data (blue) are fit to a single exponential plus a constant background (orange).}
    \label{fig:lifetime_comparison}
\end{figure}

\section{Spectral Decomposition}
While OMVPE sites have unique excitation wavelengths at the zero-phonon line, there is some overlap in the phonon-assisted excitation region included in the CEES maps in Fig. 1-2 of the main text.
In order to extract and calculate the intensities of specific sites, we used spectral decomposition using non-negative factorization to identify a components matrix and a weights matrix associated with the three strongest spectral components in the CEES maps.
This decomposition was performed using the scikit-learn Python package (sklearn.decomposition.NMF).
Initial conditions were set based on cross-sections of the CEES maps cropped to resemble the shape of known emission spectra \cite{Woodward2011SiteLayers}.
Spectral decomposition was performed to obtain the weights and components matrices for the entire CEES map for the UD sample (Fig. 2e in the main text).
For the other three samples, the components matrix was fixed to match the UD sample and only the weights were optimized.

In addition to the spectral decomposition of the CEES maps, we also used the extracted components to fit the emission spectra (Fig. 2i-l in the main text) of each sample taken under indirect excitation and extract the amplitude (area under the curve) of each component.
Fits were performed to the three spectral components, plus a constant background.
Fig. \ref{fig:spectral_decomposition_abg} shows the results of this fitting for the above-bandgap (351 nm) emission spectra, and Fig. \ref{fig:spectral_decomposition_bbg} shows the results for the just-below-bandgap (364 nm) emission spectra.
Though uncertainties are not quantified in these plots, the results should be understood with a few caveats.
These emission spectra were not taken with the intention of comparing absolute intensities, nor where their intensities corrected in the same manner as the CEES measurements were, so there may be some experimental artifacts in the intensities between samples.
While the spectral decomposition of the CEES maps used to obtain the components matrix was performed on measurements taken at 10K, the emission spectra were measured at 20K.
This temperature discrepancy reduces the quality of the fits due to broadening of the emission spectra at higher temperatures.
Finally, the fits only included contributions from the three most prominent components seen in the CEES maps, but additional spectral components may be present under indirect excitation.
With these caveats in mind, the extracted amplitudes clearly show an increased contribution of OMVPE4 in the samples with thinner active layers.

\section{Resonant Brightnesses}

The brightnesses of the different incorporation sites for each sample under resonant excitation (Fig. 3b in the main text) were obtained from the CEES maps (Fig. 2e-h) using spectral decomposition as discussed above.
These CEES maps were not originally taken with the intention of comparing absolute PL intensities, so we performed additional measurements of OMVPE4 to confirm and calibrate the amplitudes of the CEES maps.
These measurements were taken in a custom-built confocal microscope setup using a tunable OPO source (H\"{u}bner C-WAVE).
Emission from the $~^5\!D_0 \rightarrow ~^7\!F_2$ transition was isolated spectrally using a bandpass filter (FB620-10) and temporally using an accousto-optic modulator (Isomet 1250C) to pulse the excitation beam, an arbitrary waveform generator (Tektronix AWG520) to trigger the AOM and route counts from a single-photon counting module (Laser Components, Count T-100) to a data acquisition unit (National Instruments PCIe-6323) using a microwave switch (Mini-Circuits ZYSWA-2-50D).
Samples were mounted on the same puck in a closed-cycle cryostat (Montanta Instruments Cryostation) at 10K, allowing for all four samples to be measured under identical conditions.
Representative spatial photoluminescence scans are plotted in Fig. \ref{fig:pl_scans}.
The average and standard deviation of the count rates in the spatial scans were used to calibrate the intensity and uncertainty associated with the amplitude of OMVPE4 for each sample.
The intensities of each site extracted using spectral decomposition were then scaled in order to match the OMVPE4 intensity measured here.

This same setup was used to measure the temperature-dependent intensity of OMVPE4 under resonant excitation in the 10:1 DD sample, which is plotted in Fig. 5b in the main text.
In order to account for drift of the resonant frequency, coarse photoluminescence excitation scans were performed at each temperature to ensure that the excitation wavelength was well-centered on the excitation peak.

\section{Lifetime Measurements}
Lifetime measurements of OMVPE4 were measured for all four samples at 4K using the same confocal setup for measuring the resonant brightnesses.
These lifetime measurements are plotted in Fig. \ref{fig:lifetime_comparison}, with each fitted to a single exponential plus a constant background.

\bibliography{references}